\documentclass[prd,preprintnumbers,amsmath,amssymb,nofootinbib,showkeys,10pt]{revtex4-2} 

\usepackage{amsmath} 
\usepackage{amsfonts} 
\usepackage{graphicx}
\usepackage{dcolumn}
\usepackage{bm}
\usepackage{xcolor}
\usepackage{hyperref}
\usepackage{caption}
\usepackage{subcaption}
\usepackage{multirow}
\begin{document}

\title{Dark Sector Tunneling Field Potentials for a Dark Big Bang}

\author{Richard Casey, Cosmin Ilie}

\affiliation{Department of Physics and Astronomy, Colgate University, 13 Oak Drive, Hamilton, NY 13346, USA}

\begin{abstract}

All of the significant evidence for dark matter observed thus far has been through its  gravitational interactions. After 40 years of direct detection experiments, the parameter space for Weakly Interacting Massive Particles (WIMPs) as dark matter candidates is rapidly approaching the neutrino floor. Moreover, there have been no signs of WIMPs in any particle production experiments so far. In this light, we consider a dark sector that is strongly decoupled from the visible sector, interacting exclusively through gravity. In this model, proposed by Freese and Winkler~\cite{freese_dark_2023}, dark matter can be produced through a first-order phase transition in the dark sector dubbed “The Dark Big Bang.” We fully determine the allowed region of parameter space for the tunneling potential that leads to the realization of a Dark Big Bang and is consistent with all experimental bounds available. 

\end{abstract}


\maketitle 

\section{Introduction}\label{sec:Intro}

Dark matter (DM) has been researched since the 1930s and the fact that it exists in the universe is well established. The early indicators of dark matter's existence came from extragalactic data. To explain the velocities of galaxies within clusters~\cite{zwicky_1933} and the internal stability of galaxies themselves requires the presence of extra non-luminous mass~\cite{einasto_dynamic_1974,ostriker_size_1974,ostriker_Numerical_1973}. Advanced measurements of the anisotropies in the Cosmic Microwave Background (CMB) have provided the strongest evidence for the existence of dark matter. In 2018, the Planck collaboration found that 27\% of the energy of the universe is in dark matter, while only 5\% is in Standard Model (SM) particles~\cite{aghanim_2020planck}. The interested reader can refer to this comprehensive history of the discovery and ongoing search for dark matter~\cite{bertone_history_2018}. Modern research efforts aim to uncover the properties of dark matter particles through particle colliders~\cite{berlin_LHC_2019}, direct detection experiments~\cite{akerib2020LuxZep}, and observations of the universe~\cite{ade2019simons,abazajian2019cmbs4,nanograv_GWB_2023}. 

WIMPs have been one of the leading candidates for particle dark matter since those hypothetical particles are naturally produced with the observed DM relic abundance. WIMPs exist naturally in any supersymmetric extension to the standard model of particle physics. Supersymmetry (SUSY) is a theoretical model proposing the existence of an array of yet-undetected particles mirroring the standard model. It helps explain the mass value of the Higgs boson and the vacuum energy of the universe. WIMPs, being the lightest of the non-SM particles required by SUSY, should be easily detectable in most models. Yet, besides a contentious signal from the DAMA collaboration~\citep{BERNABEI:1998,Bernabei:2014,Bernabei:2018}, which may be verified or ruled-out in upcoming direct detection experiments such as COSINE-100~\citep{adhikari2019COSINE,adhikari2023induced}, no WIMPs have been conclusively detected. In fact, the parameter space above the neutrino floor becomes more and more squeezed in view of the ever increasing sensitivity of direct detection experiments~\citep[e.g.][]{cushman2013snowmass}. Moreover, no SUSY signatures have been found by the Large Hadron Collider (LHC). In this light, exploring DM models outside of the DM paradigm becomes paramount. 

Strongly decoupled dark matter only interacts with the visible sector through gravity. This is known as the ``nightmare scenario" for physicists looking for dark matter. Gravity is by far the weakest of the fundamental forces in nature, so if this is the only way dark matter speaks to the visible universe, it will be very hard to detect. The Dark Big Bang theory has been proposed in 2023 by~\citeauthor{freese_dark_2023} as an alternative to WIMPs. In this scenario, DM is generated in the early universe via a first order phase transition and is strongly decoupled. However,  if dark matter originated in a Dark Big Bang (DBB), gravity waves (GWs) produced in the phase transition could be detected, providing a source of evidence for this decoupled dark matter~\cite{freese_dark_2023}. For certain dark matter models, such as Dark WIMPs considered in this paper, evidence for dark matter produced in a DBB can be observed through correlated GW and CMB signals~\cite{freese_dark_2023}.

The Dark Big Bang model proposes that dark and visible matter could have different origins in the early universe, assuming that dark matter is strongly decoupled from the visible sector. Initially, the dark sector is cold and has energy density $\rho_{\phi}$, which is always subdominant to the energy density of the visible sector~\cite{freese_dark_2023}. Eventually, the dark false vacuum will decay, analogous to the vacuum decay of the standard Hot Big Bang scenario. The dark sector then goes through a period of reheating, producing a dark sector thermal bath denoted by $T_{\rm{DS}}$. The DBB model has been shown to successfully produce a wide range of dark matter particles with masses $\mathcal{O}$(keV - 10$^{12}$GeV)~\cite{freese_dark_2023} making it a versatile origin for many different dark matter particle models. In fact, a DBB could be possible with a dark sector that interacts weakly with the visible sector through portals such as the Higgs, neutrino, vector, or axion portal.  These DBB scenarios have yet to be researched.

By requiring that the Dark Big Bang model is consistent with standard $\Lambda$CDM cosmology and all available relevant cosmological observations, we fully determine the parameter space of the dark sector tunneling field potential. This is important, since it governs many aspects of the phase transition and thus the observable consequences (via GWs and CMB measurements) of the DBB. We will show the parameter space for the tunneling field potential has two distinct regions (see Fig.~\ref{fig: Region 2}). Parameters from one of those two have not yet been utilized or explored in the original DBB work~\cite{freese_dark_2023}, so this newly uncovered region will be of specific interest. We will also explicitly show that a DBB with parameters chosen from this new region can produce the correct relic abundance of dark matter in the universe~(see Fig.~\ref{fig: Evolution}), and lead to GWs observable with upcoming Pulsar Timing Arrays (PTAs), such as IPTA and SKA~(see Fig.~\ref{fig: GW Projections}). 

The remainder of this paper is organized as follows: in Sec.~\ref{sec:DBB} we review the main ingredients of the Dark Big Bang (DBB) model~\citep{freese_dark_2023}; in Sec.~\ref{sec:Mapping} we describe the method we use in order to constrain the tunneling potential parameters that control the DBB phenomenology; in Sec.~\ref{sec:Results} we find all  regions in the tunneling potential parameter space that lead to a DBB and are consistent with all experimental bounds; in Sec.~\ref{sec:Discussion} we discuss the implications of the regions identified. We end with conclusions in Sec.~\ref{sec: Conclusion} and two Appendices. 

\section{A Phase Transition in the Dark Sector}\label{sec:DBB}

\subsection{Background}

\begin{figure}[!htb]
 \includegraphics[width = 0.6\textwidth]{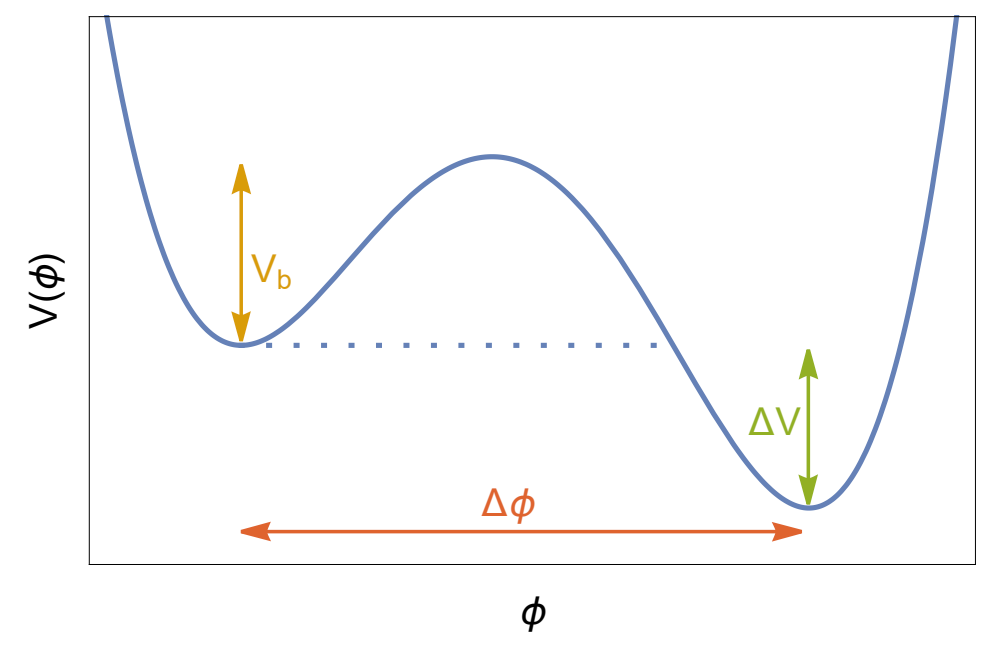}
 \caption{Model tunneling field potential used for a first-order phase transition in the dark sector (figure reproduced from~\cite{freese_dark_2023}). At the beginning of the universe, $\phi$ sits in the metastable minimum of the potential well. The discontinuous phase transition is realized when $\phi$ quantum tunnels through the potential barrier and eventually comes to rest in the ultrastable minimum. The height of the potential barrier is denoted by $V_{\rm{b}}$. The separation between the two minima is $\Delta\phi$. The change in potential of $\phi$ through the phase transition is $\Delta V$.}
 \label{fig:DBB Potential}
 \end{figure}

The Dark Big Bang occurs through a first-order phase transition in the dark sector.\footnote{Other models considering higher-order phase transitions can be studied in the future~\cite{freese_dark_2023}.} Since the dark sector is initially cold, the phase transition occurs through the quantum tunneling of the $\phi$ field through a potential barrier (as opposed to thermal fluctuations causing $\phi$ to jump over the potential barrier). Initially, all of the dark sector energy density resides in the tunneling field $\phi$. This phase transition can be modeled like a particle in a potential with two non-degenerate minima. An illustrative example is shown in Fig.\ref{fig:DBB Potential} (reproduced from~\cite{freese_dark_2023}).

While the dark sector is in the false vacuum state ($\phi$ sits in the metastable minimum of $V(\phi)$), quantum fluctuations cause bubbles of the true vacuum to randomly nucleate throughout the universe. If the bubbles nucleate with a large enough radius, expansion will be energetically favored over collapse and the bubbles will expand and collide~\cite{coleman1977fate}. In Fig.~\ref{fig:DBB Potential}, this can be modelled by the $\phi$ field quantum tunneling through the potential barrier and materializing at a height $V(\phi)=\Delta V$ near the ultrastable minimum with zero kinetic energy. Once tunneled, the field $\phi$ rolls towards the ultrastable minimum, with its gained kinetic energy stored in the expanding bubble walls~\cite{coleman1977fate}. $\phi$ will then undergo oscillations around the ultrastable minimum, which are dampened by couplings to dark sector fields. As $\phi$ comes to a rest in the true vacuum state, bubble walls collide, releasing their energy into dark sector fields and gravity waves. A first-order phase transition is phenomenologically similar to the phase transition of super-heated water to gas~\cite{coleman1977fate}: bubbles of the new phase (gas) nucleate and expand in the old phase (water) until all of the water has been turned into gas.

The processes described above are controlled by the Lagrangian for the dark sector. Following~\cite{freese_dark_2023}, this is

\begin{equation}
    \mathcal{L}_{\rm{DS}}=\frac{1}{2}\partial_{\mu}\phi\partial^{\mu}\phi+\frac{1}{2}\partial_{\mu}\chi\partial^{\mu}\chi - V(\phi)-y\phi^{2}\chi^{2}-\frac{m_{\chi}^{2}}{2}\chi^{2}-\kappa\chi^{4} 
      \hspace{2 mm} (+\mathcal{L}_{\rm{DR}})
    \label{eq:Dark Sector Lagrangian}
\end{equation}
where both $\phi$ and $\chi$ are taken to be real scalars, and $\mathcal{L}_{\rm{DR}}$ is the dark radiation Lagrangian density, if dark radiation is included in the model being considered. The tunneling field potential $V(\phi)$ takes the form~\cite{freese_dark_2023}

\begin{equation}
    V(\phi) = \frac{m^{2}}{2}\phi^{2}-\mu\phi^{3}+\lambda\phi^{4}+\Delta V
    \label{eq:Tunneling Potential}
\end{equation}
to allow for two non-degenerate potential minima. Placing the metastable minimum at $\phi = 0$ requires $\mu>\sqrt{2\lambda}m$, where $m$ is the mass of of the tunneling field in the false vacuum state and $\mu$, $\lambda$ govern $\phi$ interactions. The true minimum of $V(\phi)$ is then~\cite{freese_dark_2023}

\begin{equation}
    \phi_{min} = \frac{3\mu+\sqrt{9\mu^{2}-16\lambda m^{2}}}{8\lambda} = \Delta\phi
    \label{eq:Potential minima}
\end{equation}
which is found by setting $V'(\phi) = 0$ and taking the positive root. Since $\Delta V = V(0) - V(\phi_{\rm{min}})$, $\Delta V$ can be solved for algebraically~\cite{freese_dark_2023}

\begin{equation}
    \Delta V = \frac{(3\mu+\sqrt{9\mu^{2}-16\lambda m^{2}})^2}{64\lambda^{2}}\left(\frac{3\mu^{2}+\mu\sqrt{9\mu^{2}-16\lambda m^{2}}-8\lambda m^{2}}{32\lambda}\right)
    \label{eq: Delta V}
\end{equation}

Using a similar approach, the barrier height is

\begin{equation}
   V_{b} = \frac{(3\mu-\sqrt{9\mu^{2}-16\lambda m^{2}})^2}{64\lambda^{2}}\left(\frac{-3\mu^{2}+\mu\sqrt{9\mu^{2}-16\lambda m^{2}}+8\lambda m^{2}}{32\lambda}\right)
    \label{eq: Barrier Height}
\end{equation}

 The $\phi$ field undergoes a mass shift during the phase transition. When $\phi$ is in the true vacuum state, its mass becomes~\cite{freese_dark_2023}

\begin{equation}
    m_{\phi}^{2} = V''(\Delta\phi)=\frac{9\mu^{2}+3\mu\sqrt{9\mu^{2}-16\lambda m^{2}}}{8\lambda}-2m^2
    \label{eq:m phi}
\end{equation}
which can be seen through the MacLaurin Series expansion of $V(\phi)$ around the metastable minimum.

\subsection{Phase Transition Parameters}

Many of the physical characteristics of a DBB are determined by the shape and scale of the tunneling potential $V(\phi)$. This paper will focus on the two parameters that are constrained by cosmological observations: the visible sector temperature at the time of the DBB ($T_{*}$) and the strength of the DBB ($\alpha$). The strength of the DBB is defined as the ratio of dark to visible sector energy densities directly after the phase transition~\cite{freese_dark_2023}. Since the DBB occurs early in the universe during the radiation-dominated epoch, the total energy density of the visible sector can be approximated as the energy stored in radiation. The strength of the DBB can be written as~\cite{freese_dark_2023} 

\begin{equation}
    \alpha \simeq \frac{\rho_{\phi}}{\rho_{r,*}} \simeq \frac{\Delta V}{\frac{\pi^2}{30}g_{\rm{eff}}(T_{*})T_{*}^4}
    \label{eq: alpha definition}
\end{equation}

The $_{*}$ subscript notation means the variable is evaluated at the time of the DBB. We have also defined the energy density of the dark sector before the phase transition as $\rho_{\rm{DS}}\simeq\rho_{\phi} = \Delta V$~\cite{freese_dark_2023}. The DBB theory accounts for the creation of all dark matter in the decay of the dark false vacuum, so dark matter particles must be produced efficiently in this phase transition. This requires the presence of light (relativistic) degrees of freedom in the dark sector $g_{\rm{DS}}$ that the bubble walls can always efficiently decay into~\cite{freese_dark_2023}. With these light degrees of freedom in the dark sector, the end of the phase transition is considered a period of dark sector reheating, and the dark plasma produced quickly reaches a thermal equilibrium described by $T_{\rm{DS}}$~\cite{freese_dark_2023}. Gravity waves make up at most 5\% of the dark sector energy density released in the phase transition~\cite{freese_dark_2023}, so the energy density in the dark sector after the phase transition can be approximated by~\cite{freese_dark_2023}

\begin{equation}
    \rho_{\phi} = \Delta V \simeq \frac{\pi^2}{30}g_{\rm{DS}}(T_{\rm{DS,*}})T_{\rm{DS,*}}^4
    \label{eq: DS energy density}
\end{equation}

With this approximation, the dark reheating temperature can then be defined as~\cite{freese_dark_2023}

\begin{equation}
    T_{\rm{DS,*}} \simeq \left(\frac{30}{\pi^2g_{\rm{DS,*}}}\Delta V\right)^{1/4}
    \label{eq: Tds def}
\end{equation}
where $g_{\rm{DS,*}} \equiv g_{\rm{DS}}(T_{\rm{DS,*}})$.
Then $\alpha$ can be written as the ratio of the dark sector reheating temperature to the visible sector temperature when the DBB occurs, up to their respective degrees of freedom

\begin{equation}
    \alpha \approx \frac{g_{\rm{DS},*}}{g_{\rm{eff},*}}\left(\frac{T_{\rm{DS},*}}{T_{*}}\right)^4
    \label{eq: alpha simplified}
\end{equation}

\subsection{Bounds From the Early Universe}
\label{subsec: Bounds from early universe}

The DBB is bounded to be consistent with cosmological observations and standard $\Lambda$CDM cosmology. The bounds are discussed at length in Ref.~\cite{freese_dark_2023} and we will provide a brief overview here. The lower limit on $\alpha$ is determined by the requirement that the energy released in the phase transition can account for all dark matter in the universe today~\cite{freese_dark_2023}. In mapping the power spectrum of CMB anisotropies, the Planck collaboration determined today's dark matter density to be $\rho_{\rm{DM},0} = 1.26$ keVcm$^{-3}$ ~\cite{aghanim_2020planck}. Requiring that all of the DM today was produced in a DBB leads to the following lower bound on the strength of the DBB (see~\cite{freese_dark_2023})

\begin{equation}
    \alpha \geq \frac{\rho_{\rm{DM},0}}{a_{*}^3\rho_{r,*}} = \frac{4}{3}\frac{\rho_{\rm{DM},0}}{s_{0}T_{*}} = 5.8\times10^{-7}\left(\frac{\rm{MeV}}{T_{*}}\right)
    \label{eq: alpha lower bound}
\end{equation}
where radiation density has been put in terms of entropy density of the visible sector $\rho_{r,*} = (3/4)s_{*}T_{*}$ and conservation of entropy $s_{*}a_{*}^3 = s_{0}$ has been utilized. 

The upper limit of $\alpha$ comes from existing bounds on the amount of extra energy density present in the early universe (in the DBB scenario, that extra energy density is stored in $\rho_{\rm{DS}}$)~\cite{freese_dark_2023}. Extra energy density causes a speed up of the expansion of the universe realized as an increase in the Hubble parameter $H$. The speed up could cause weak interactions to freeze out earlier in the universe, leading to an increase in the relativistic species populating the early universe $N_{\rm{eff}}$. $N_{\rm{eff}}$ counts the number of these relativistic species (not including photons), and is called the effective number of neutrinos (although particles other than neutrinos can contribute to $N_{\rm{eff}}$)~\cite{baumann_cosmo_2022}. An increase in this number leads to dampening in the small scale (large $\ell$) power spectrum of CMB anisotropies. The Standard Model predicts $N_{\rm{eff}} = 3.046$~\cite{baumann_cosmo_2022} and the Planck collaboration has determined bounds on the change in this number $\Delta N_{\rm{eff}} = 0.22\pm 0.15$~\cite{aghanim_2020planck}. This suggests $\Delta N_{\rm{eff}} < 0.5$, which leads to the following upper limit on $\alpha$~\cite{freese_dark_2023}

\begin{equation}
    \alpha < \left(\frac{2\pi^2}{45}\right)^{1/3}\frac{4g_{\rm{eff}}^{1/3}(T_{*})\rho_{\rm{DR},0}}{3s_{0}^{4/3}}< 0.079\left(\frac{g_{\rm{eff}}(T_{*})}{10}\right)^{1/3}
    \label{eq:alpha upper bound}
\end{equation}
where $\rho_{\rm{DR,0}}$ is the present dark radiation density $\rho_{\rm{DR,0}} \lesssim 29.6$ meVcm$^{-3}$~\cite{freese_dark_2023}. 

The final constraint is on the latest a DBB can occur to be consistent with the $\Lambda$CDM picture of structure formation. Requiring that dark matter obtains the correct adiabatic perturbations to seed structure formation leads to the strongest constraint on the latest a DBB can occur~\cite{freese_dark_2023}. $\rho_{\phi}$ is a smooth fluid at the time of the DBB, so the dark matter content must pick up perturbations from the visible sector radiation bath which dominates the universe~\cite{freese_dark_2023}. The radiation bath exhibits adiabatic perturbations, which are realized in a common, local time shift of all background quantities in cosmological perturbation theory~\cite{baumann_cosmo_2022}. Adiabatic perturbations are defined by the equality $\delta\rho_{r}(\eta,\textbf{x}) = \bar{\rho_{r}}'\pi(\eta,\textbf{x})$ where $\eta$ is conformal time and \textbf{x} is a three dimensional space coordinate. This equation says that perturbed quantities ($\delta\rho$) at some spacetime point are the same as those in the background universe ($\bar{\rho}$) at a slightly shifted spacetime point. Adiabatic perturbations in visible sector radiation cause the DBB phase transition to occur at slightly different times throughout the universe. These time differences result in the perturbations becoming imprinted on the dark matter produced in the DBB (see~\cite{freese_dark_2023}).

Observations of Ly-$\alpha$ from distant quasars can be used to constrain these perturbations on large scales. Using existing constraints on cosmological fluctuations from warm dark matter simulations~\cite{garzilli2021WDM,villasenor2023WDMLyalpha2}, the approximate bound on the latest a DBB can occur is~\cite{freese_dark_2023}
    
\begin{equation}
    T_{*} > 6.8 \times 10^{-4} \rm{MeV}
    \label{eq: T star bound}
\end{equation}

\begin{figure}[!htb]
 \includegraphics[width = 0.8\textwidth]{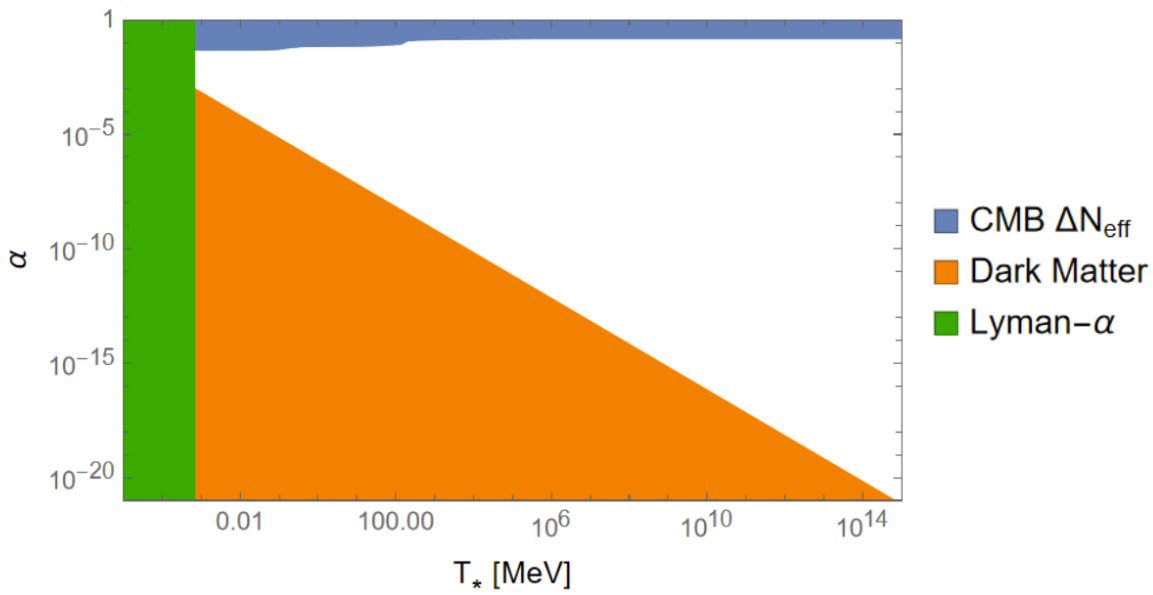}
 \caption{Bounds on the strength of a DBB and the latest a DBB can occur~\cite{freese_dark_2023}. The allowed region for a DBB consistent with observations is the white space in the figure. The upper bound on $\alpha$ depends on the relativistic degrees of freedom in the universe and remains constant above the mass scale of the heaviest SM particle: top quarks m$_{t}\sim 1.8\times10^{5}$MeV. The lower bound on $\alpha$ continues to decrease linearly with temperature. The Ly-$\alpha$ bound is constant and is a bound on the latest time in the universe a DBB can occur.}
 \label{fig:DBB Bounds}
 \end{figure}

The bounds discussed above are summarized in Fig.~\ref{fig:DBB Bounds}, which has been expanded from~\cite{freese_dark_2023} to include higher temperatures. The first aim of our work is to map this allowed parameter space (white gap region in Fig.~\ref{fig:DBB Bounds}) onto constraints on the parameters determining the tunneling field potential, which in turn control many aspects of a DBB model. In the next section, we will show how the phase transition parameters can be recast as functions of the tunneling field parameters.

\section{Mapping the Bounds On to tunneling Field Parameters}\label{sec:Mapping}

The initial conditions of a DBB are determined by the shape and scale of the tunneling potential (Eq.~\ref{eq:Tunneling Potential}). The input parameters for a DBB model are then $m$, $\mu$, and $\lambda$, as well as dark matter particle mass and couplings, which are model dependent. The physical characteristics of the phase transition can be determined directly from the input parameters, as will be shown in the following section.

\subsection{Bubble Nucleation and Decay}

As outlined above, the DBB takes place through a first-order phase transition in the dark sector in which bubbles of true vacuum nucleate, expand, and collide, transforming the false vacuum into true vacuum. The process of false vacuum decay through quantum tunneling has been studied in the context of the Hot Big Bang~\cite{coleman1977fate,callan1977fate2} and the developed formalism can be applied to the DBB. The bubble nucleation rate per comoving volume is~\cite{coleman1977fate, callan1977fate2}

\begin{equation}
    \Gamma \simeq m^4\left(\frac{S}{2\pi}\right)^2 e^{-S}
    \label{eq:nucleation rat}
\end{equation}
where S is the Euclidean action of the bounce solution. The bounce solution comes from solving the Euclidean, or imaginary time ($t_{\rm{E}} = it$), equation of motion for $\phi$: $\Box_{\rm{E}} - V'(\phi) = 0$ with the boundary conditions $d\phi(0,\textbf{x})/dt_{E} = 0$ and $\phi(\pm\infty,\textbf{x}) = 0$~\cite{kolbandturner}. In Euclidean time, $\phi$ starts in the metastable minimum at $\rm{t_{E}} = -\infty$, comes to rest near the ultrastable minimum at $\rm{t_{E}} = 0$, and returns back to the metastable minimum at $\rm{t_{E}} = \infty$, hence the ``bounce". The action for the quartic potential considered here has a numerical solution beyond the analytic thin wall ($\Delta V \ll V_{b}$) approximation~\cite{adams1993bounce}

\begin{equation}
    S \simeq \frac{\pi^2\mu^6}{24\lambda(\mu^2 - 2\lambda m^2)^3}\left[13.832 \frac{4\lambda m^2}{\mu^2} -10.819 \left(\frac{4\lambda m^2}{\mu^2}\right)^2 + 2.0765\left(\frac{4\lambda m^2}{\mu^2}\right)^3\right]
    \label{eq: Euclidean action}
\end{equation}

The lifetime of the dark sector false vacuum is now defined as the time the DBB occurs~\cite{freese2022firstorderGW,freese_dark_2023}

\begin{equation}
    t_{*} \simeq \left(\frac{105}{8\pi\Gamma}\right)^{1/4}
    \label{eq: time of DBB}
\end{equation}

Using the definition of $\rho_{r}$ stated in Eq.~\ref{eq: alpha definition} and the fact that during the radiation dominated epoch $\rho_{r} \sim a^{-4} \sim t^{-2}$, the visible sector temperature at the time of the DBB can be put in terms of t$_{*}$~\cite{freese_dark_2023}

\begin{equation}
    T_{*} \simeq \left(\frac{45}{2\pi^2}\right)^{1/4}\left(\frac{M_{P}^2}{g_{\rm{eff}}(T_{*})t_{*}^2}\right)^{1/4}
    \label{eq: Tstar}
\end{equation}
where $M_{P} = 2.44\times10^{21}$ MeV is the reduced Planck mass. Additional phase transition (PT) parameters such as the duration of the PT and the initial radius of the nucleated bubbles are shown in~\cite{freese_dark_2023} but are not considered in this work. 

\subsection{Mapping Equations}
\label{subsec: Mapping Eqs}

In order to map the parameter space for a DBB on to the input parameters for a tunneling field potential, the phase transition parameters need to be put in terms of $m$, $\mu$, and $\lambda$. First we will deal with $T_{*}$. One complication that is immediately apparent from Eq.~\ref{eq: Tstar} is the dependence on $g_{\rm{eff}}$, which also depends on $T_{*}$.\footnote{This is only an issue when $g_{\rm{eff}}$ is a dynamic variable (below $T\sim 1.8\times 10^5$MeV).} Combining Eqs.~\ref{eq: Tstar},~\ref{eq: time of DBB},~\ref{eq:nucleation rat}, we find

\begin{equation}
    T_{*}g_{\rm{eff}}^{1/4}(m,\mu,\lambda) \simeq \left(\frac{45M_{P}^2}{2\pi^2}\right)^{1/4}\left(\frac{8\pi}{105}\right)^{1/8}\left(m^4\left(\frac{S(m,\mu,\lambda)}{2\pi}\right)^2 e^{-S(m,\mu,\lambda)}\right)^{1/8}
    \label{eq: Tgmmu}
\end{equation}

Note that for any given value of $T_{*}g_{\rm{eff}}^{1/4}$ one can uniquely solve for $T_{*}$, and consequently for $g_{\rm{eff}}(T_*)$. It is a simpler process putting $\alpha$ in terms of tunneling field parameters, since the relativistic degrees of freedom in both the visible and dark sectors cancel out in the expansion. From Eqs.~\ref{eq: alpha definition} and~\ref{eq: Tgmmu}, $\alpha$ becomes

\begin{equation}
    \alpha(m,\mu,\lambda) \simeq \frac{60}{45M_{P}^2}\left(\frac{105}{8\pi}\right)^{1/2}\left(m^4\left(\frac{S(m,\mu,\lambda)}{2\pi}\right)^2 e^{-S(m,\mu,\lambda)}\right)^{-1/2}\Delta V(m,\mu,\lambda)
    \label{eq: alpha mmu}
\end{equation}
where $S$ and $\Delta V$ are defined in Eqs.~\ref{eq: Euclidean action} and~\ref{eq: Delta V}. With these equations, we can recast the bounds described in Sec.~\ref{subsec: Bounds from early universe} entirely as functions of $m$, $\mu$, and $\lambda$. 

\section{Results}\label{sec:Results}

\begin{figure}[!htb]
 \includegraphics[width = 0.8\textwidth]{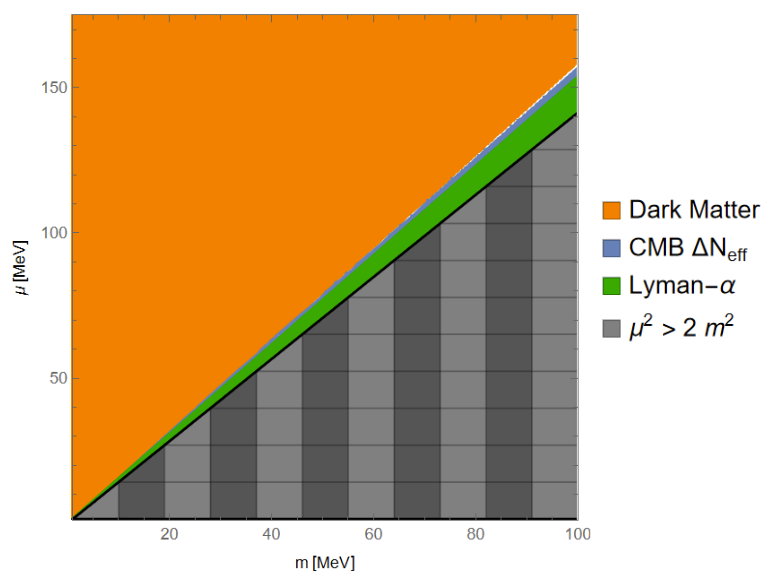}
 \caption{The available parameter space for a DBB when $\mu \gtrsim  m$ (Region 1). The bounds have been recast from Fig.~\ref{fig:DBB Bounds} to $m$-$\mu$ space using Eqs.~\ref{eq: Tgmmu} and~\ref{eq: alpha mmu}. Also plotted is the constraint that $\mu > \sqrt{2\lambda m}$ with $\lambda=1$. The white region shows tunneling field parameter choices for a DBB consistent with $\Lambda$CDM and widens (narrows) beyond the figure to the right (left). While it appears that the parameter space is closed for most of the figure, Fig.~\ref{fig: Sliver Zoom 4} will show that the parameter space remains open throughout this range and is too narrow to visualize at this scale.}
 \label{fig:Sliver Region Full}
 \end{figure}

\begin{figure}[!htb]
 \includegraphics[width = 0.9\textwidth]{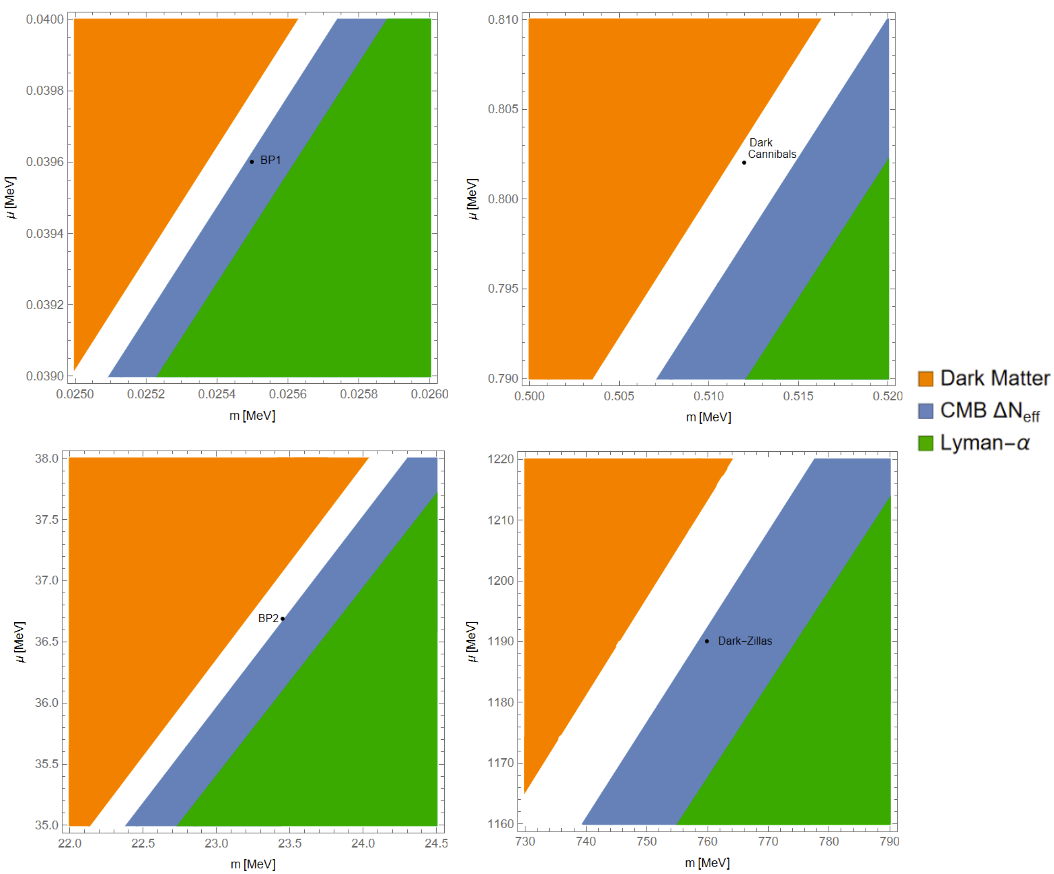}
 \caption{Four different magnifications into Region I of the parameter space, which was partially shown Fig.~\ref{fig:Sliver Region Full}. The regions chosen correspond to the four benchmark Dark Big Bangs considered in~\cite{freese_dark_2023}. Interestingly, we found that the choices of parameters for BP1 (Panel 1) and Dark-Zillas (Panel 4) slightly violate the upper bound on $\alpha$. These discrepancies do not significantly impact the results of~\cite{freese_dark_2023}, as the parameters can be adjusted slightly to produce the same phase transition characteristics used in their analysis.}
 \label{fig: Sliver Zoom 4}
 \end{figure}

\begin{figure}[!htb]
 \includegraphics[width = 0.8\textwidth]{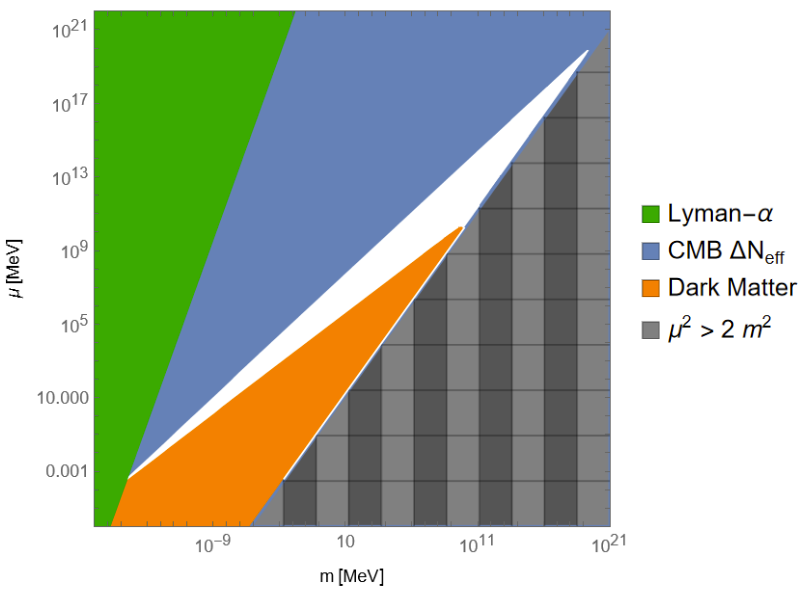}
 \caption{Full parameter space for tunneling potential parameters that lead to a DBB consistent with observations. At low $m$ and $\mu$ values, Region 1 and and Region 2 are separated by the lower bound on $\alpha$ (i.e. orange excluded region). On this log scale, Region 1 is barely visible, yet remains open and is represented by the white line just above the $\mu^2 > 2m^2$ region. For $m \gtrsim 6.1\times10^{9}$, the lower (orange) bound on $\alpha$ plays no role and the parameter space is bounded on both sides by the upper (blue) bound.}
 \label{fig: Region 2}
 \end{figure}

For our analysis, we have fixed $\lambda$ = 1 to allow for the decay of $\phi$ to be very efficient (this choice was made for each of the benchmark Dark Big Bangs considered in~\cite{freese_dark_2023}). Future research could look into how the parameter space changes with $\lambda$. For reasons outlined in Appendix~\ref{append: Analytic Behavior}, we will discuss the parameter space in two regions. Region 1, which we have dubbed ``the sliver", rides the singularity in $S$ (where $\mu^2 = 2\lambda m^2$). This region's proximity to the singularity makes the phase transition (PT) parameters very sensitive to choices of $m$ and $\mu$. Furthermore, this region of parameter space is so narrow that it cannot be visualized over large ranges of $m$ and $\mu$. We find that a second region of parameter space (Region 2) opens up when $\mu \gg$ $m$ (see Appendix~\ref{append: Analytic Behavior}). Region 1 of the available parameter space for a DBB is shown in Figs.~\ref{fig:Sliver Region Full} and~\ref{fig: Sliver Zoom 4}. Both regions are shown in Figs.~\ref{fig: Region 2} and~\ref{fig: All region 2 Bounds}.

Throughout much of the parameter space, the bound on the latest a DBB can occur (Lyman-$\alpha$ bound, e.g. green swath) is weaker than the upper (CMB $\Delta N_{\rm{eff}}$, e.g. blue  excluded region) and lower (Dark Matter, e.g. orange excluded region) bounds on the strength of the DBB $\alpha$. However, this bound does play a role at small $m$ and $\mu$ as shown in Fig.~\ref{fig: Region 2}. The Lyman-$\alpha$ bound can be turned into a constraint between $\alpha$ and $\Delta$V by combining Eqs.~\ref{eq: alpha definition} and~\ref{eq: T star bound}: $\alpha \lesssim 4.24\times 10^{12} \Delta V$ MeV$^{-4}$, where we used $g_{\rm{eff}}(6.8\times 10^{-4}$~MeV) $=3.35$. Using this constraint and Eq.~\ref{eq: alpha lower bound}, the parameter space is ruled out towards small $m$ and $\mu$ values when the Lyman-$\alpha$ constraint is equal to the Dark Matter bound

\begin{equation}
    4.24\times10^{12}\left(\frac{\Delta V(m,\mu)}{\rm{MeV}^4}\right) \simeq 5.8\times10^{-7}\left(\frac{\rm{MeV}}{T_{*}(m,\mu)}\right) = \alpha(m,\mu)
    \label{eq: T intersection}
\end{equation}
which can be solved for $m$ and $\mu$ numerically (we have set $\lambda = 1$). The values of $m$ and $\mu$ below which the parameter space is ruled out by the temperature bound are shown in the top two rows of Tab.~\ref{tab: critical points}.


\begin{figure}[!htb]
 \includegraphics[width = 1\textwidth]{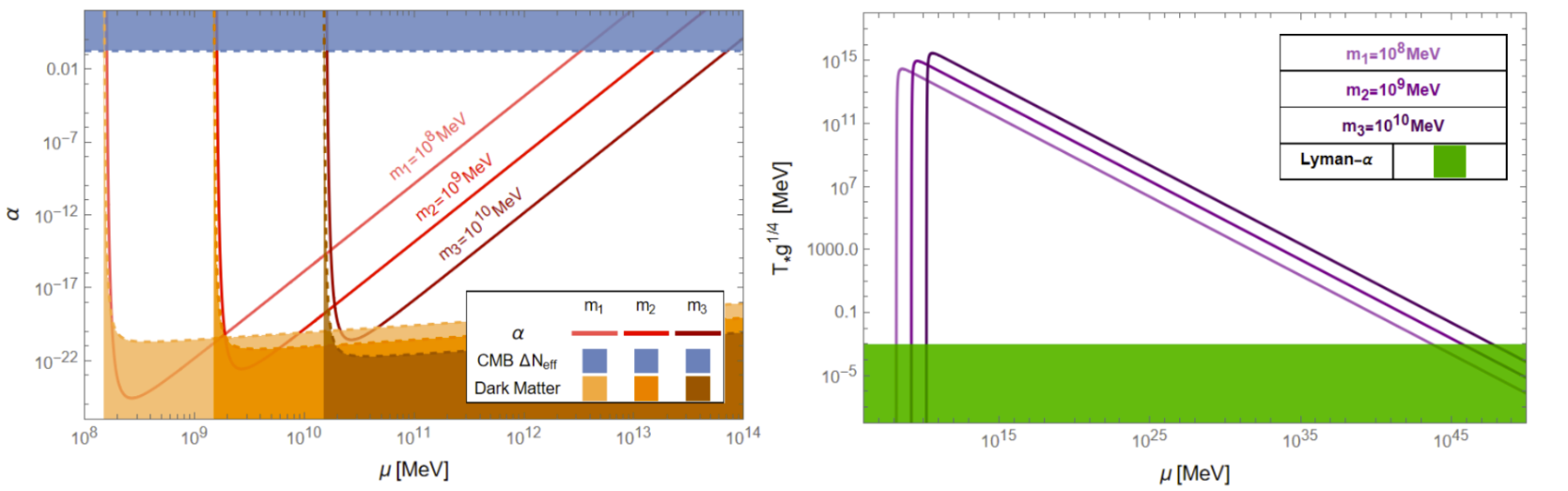}
 \caption{Left Panel: Values of $\alpha$ for fixed $m$. As before, $\alpha$ is bounded above by the CMB $\Delta N_{\rm{eff}}$ (blue) bound and below by the Dark Matter (orange) bounds. For the three $m$ values shown, the CMB $\Delta N_{\rm{eff}}$ bound is the same and constant. The Dark Matter bound changes for each fixed $m$ value. All values of $\alpha$ for the $m_{3}$ plot are above the lower bound. Right Panel: Temperature of the visible sector at the time of the DBB as a function of $\mu$. We have included the effective relativistic degrees of freedom $g_{\rm{eff}}(T)$ in our definition of temperature as in Sec.~\ref{subsec: Mapping Eqs}. Dark Big Bangs that occur at a temperature below the Lyman-$\alpha$ bound are excluded by the current understanding of structure formation.}
 \label{fig: Constant m}
 \end{figure}

 In Fig.~\ref{fig: Region 2} we identify the part of the ($\mu$ vs. $m$) parameter space that leads to a DBB which is consistent with all experimental bounds. Note that the white region (allowed) can be naturally split into two. First, the very narrow sliver (Region~1) riding along the $\mu^2=2m^2$ line, and open from $\mu\sim 0.001$ to $\mu\sim 10^{10}$. This region was previously explored in~\cite{freese_dark_2023}. Note a second, much wider allowed region (Region~2). The existence of Region~2, which was not explored in~\cite{freese_dark_2023}, is due to the value of $\alpha$ at constant $m$ increasing back through the bounds as $\mu$ increases (see Appendix~\ref{append: Analytic Behavior} for details). This behavior can be seen in Fig.~\ref{fig: Constant m}, which shows plots of $\alpha$ and $T_{*}g_{\rm{eff}}^{1/4}$ (Eq.~\ref{eq: alpha mmu} and Eq.~\ref{eq: Tgmmu}) for fixed $m$ (taking vertical slices of Figs.~\ref{fig:Sliver Region Full},~\ref{fig: Sliver Zoom 4},~\ref{fig: Region 2}). We define Region 2 as the available parameter space when $\mu \gg m$ (corresponding to the portions of the plots in Fig.~\ref{fig: Constant m} with gentler slopes). As both $m$ and $\mu$ increase, the minimum value of $\alpha$ increases, eventually evading the lower bound ($m_{3}$ in the left panel of Fig.~\ref{fig: Constant m}). For $m$ and $\mu$ greater than this point (listed in row 3 in Tab.~\ref{tab: critical points}), the parameter space is bounded on both sides by the upper bound on $\alpha$ (blue region). The parameter space closes when the minimum value of $\alpha$ is larger than the upper bound. The critical points, defining where regions of the parameter space open or close, are summarized in Tab.~\ref{tab: critical points}.

In comparing Figs.~\ref{fig:Sliver Region Full} and~\ref{fig: Sliver Zoom 4} with Fig.~\ref{fig: Region 2}, there is a change in the hierarchy of the bounds on a DBB. Namely, for Region~1 the CMB $\Delta N_{\rm{eff}}$ bound (blue) rules out larger ($m,\mu$) values than the Lyman-$\alpha$ bound (green), whereas for Region~2, the situation is reversed for most of the parameter space. The hierarchy flip is explained by Fig.~\ref{fig: Constant m}. For each $m$ value plotted in Fig.~\ref{fig: Constant m}, when $\mu \approx m$, $\alpha$ violates the CMB $\Delta N_{\rm{eff}}$ bound and $T_{*}g_{\rm{eff}}^{1/4}$ violates the Lyman-$\alpha$ bound. Increasing $\mu$, $T_{*}g_{\rm{eff}}^{1/4}$ increases into the allowed region, quickly followed by $\alpha$ decreasing into the allowed region. This behavior is shown in Figs.~\ref{fig:Sliver Region Full} and~\ref{fig: Sliver Zoom 4} where the Lyman-$\alpha$ bound is weaker than the CMB $\Delta N_{\rm{eff}}$ bound throughout Region 1. In Fig.~\ref{fig: Region 2}, the corresponding section of parameter space follows the $\mu^2 > 2m^2$ boundary (the sliver). Looking closely, the CMB $\Delta N_{\rm{eff}}$ bound can be seen (the blue line along the gray shaded region). We have chosen not to plot the Lyman-$\alpha$ bound along the gray region (as it is shown in Figs.~\ref{fig:Sliver Region Full} and~\ref{fig: Sliver Zoom 4}) because it is weaker than the blue CMB $\Delta N_{\rm{eff}}$ bound and obscures the edge of the parameter space. Going back to Fig.~\ref{fig: Constant m}, as $\mu$ continues to increase, $\alpha$ returns to violate the CMB $\Delta N_{\rm{eff}}$ bound long before $T_{*}g_{\rm{eff}}^{1/4}$ returns to violate the Lyman-$\alpha$ bound. Hence the hierarchy flip of the bounds as we move vertically from below Region 1 to above Region 2.

 \begin{table}[!htb]
 \centering
    \begin{tabular}{|c|c|}\hline
     Critical Point  & ($m, \mu$) [MeV] \\ \hline
     Region 2 opens & ($1.9\times10^{-16},2.1\times10^{-4}$) \\
     Region 1 opens & $(2.3160\times10^{-4}, 3.5915\times10^{-4})$ \\
     Regions merge & ($6.10\times10^9, 1.64\times10^{10}$)\\
     Parameter space closes (TCC) & ($1.65\times10^{12},2.70\times10^{12}$)\\
     Parameter space closes & ($8.17\times10^{19}, 2.18\times10^{20}$) \\ \hline
    \end{tabular}
    \caption{Numerical solutions for the critical points of the parameter space in order of increasing $m$. The parameter space is closed below the ``opening" values and above the ``closing" values. The two different values for where the parameter space closes are from whether or not the Trans-Planckian Censorship Conjecture (TCC) is considered as a bound on the scale of inflation (see Sec.~\ref{sec:Discussion}). The opening point of Region 1 must be approximated to smaller orders of magnitude because of the sensitivity of $\alpha$ near the singularity.}
    \label{tab: critical points}
\end{table}

\section{Discussion}\label{sec:Discussion}

We are particularly interested in looking at the qualitative differences between the two regions. In Region 2, defined by $\mu \gg m$, analytic analysis is much simpler than in the sliver (Region~1). Additionally, there has yet to be a benchmark DBB considered with input parameters drawn from this region. Given these considerations, this section will first look at the features of tunneling potentials from Region 2 and then consider benchmark DBBs using input parameters from this region.

\subsection{Tunneling Potentials}\label{subsec:Tunneling potentials}

\begin{figure}[!htb]
 \includegraphics[width = 0.7\textwidth]{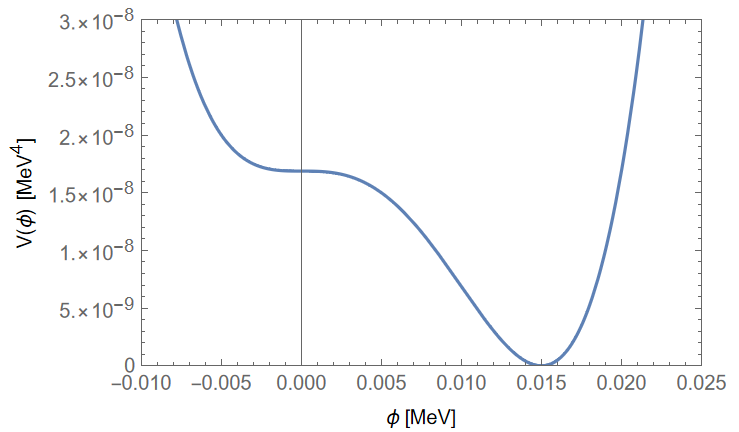}
 \caption{Benchmark tunneling field potential from Region 2 of the parameter space. The potential shown here corresponds to the input parameters of DW 1 shown in Tab.~\ref{tab: DBB Benchmarks}. Although the potential barrier is imperceptible, it is large enough to withstand thermal fluctuations. For this potential, $V_{b}\sim 10^{-77}$ MeV$^4$ while $T_{\rm{GH}}^{4}\sim 10^{-106}$ MeV$^4$.}
 \label{fig: DW1 Potential}
 \end{figure}

Looking at the quartic potential in Eq.~\ref{eq:Tunneling Potential}, one can deduce the effects $m, \mu$, and $\lambda$ have on the shape of the potential. As $\phi$ increases from zero, $V(\phi)$ first increases $\sim m^2\phi^2$, then decreases $\sim \mu\phi^3$, and finally increases $\sim\lambda\phi^4$. So throughout Region 2, where $\mu \gg m$, the potential barrier protecting the stability of the false vacuum is the smallest feature of the tunneling potential, as shown in Fig.~\ref{fig: DW1 Potential}. This section will show that although $V_{b}$ is small throughout Region 2, the tunneling potentials from this region can successfully provide the initial conditions for a first-order phase transition in the dark sector.

Additional constraints on the tunneling field potential come from the stability of the dark false vacuum. While the tunneling field is trapped in the metastable minimum of the potential, the dark sector goes through a period of inflation. For the dark false vacuum to remain stable during inflation, the potential barrier ($V_b$) must exceed fluctuations caused by the Gibbons-Hawking temperature $T_{\rm{GH}}$~\cite{freese_dark_2023,hawking1982supercooled}
\begin{equation}\label{eq:GHbound}
V_{b}\gtrsim(T_{\rm{GH}})^4 .   
\end{equation}
The Gibbons-Hawking temperature is proportional to the Hubble parameter during inflation $T_{\rm{GH}} = H_{I}/2\pi$~\cite{gibbons-hawking1977}. We then have the condition\footnote{A weaker constraint on the stability of the false vacuum $\Delta V \lesssim 3M_{p}^2T_{*}^2$ is considered in~\cite{freese_dark_2023}.}

\begin{equation}
    V_{b}\gtrsim \left(\frac{1}{2\pi}\right)^{4}\left(\frac{\Delta V}{3 M_{P}^2}\right)^{2}
    \label{eq: potential barrier bound}
\end{equation}
where the Hubble parameter of the dark sector during inflation is defined with $\rho_{\rm{DS}} = \Delta V$. In Region 2, we can safely use analytic approximations for $V_{b}$ and $\Delta V$. It can be shown that for $\mu \gg \rm{m}$

\begin{equation}
    \Delta V \approx \frac{27}{256}\mu^4\left(1- \frac{4}{3}\frac{m^2}{\mu^2}\right)
    \label{eq: Delta V approx}
\end{equation}
and 
\begin{equation}
    V_{b} \approx \frac{576m^6}{31104\mu^2}
    \label{eq: Vb approx}
\end{equation}

\begin{figure}[!htb]
 \includegraphics[width = 0.8\textwidth]{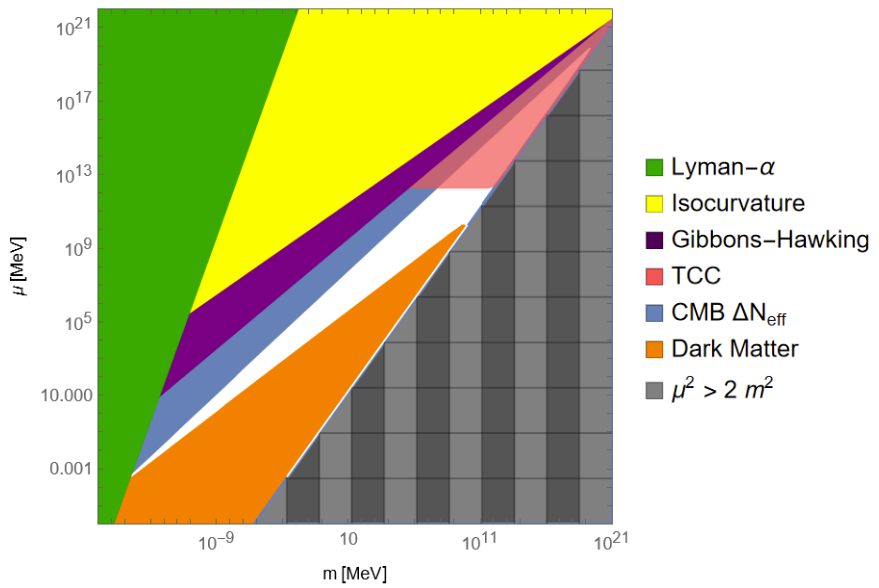}
 \caption{The available parameter space for a DBB consistent with observations, including constraints on the scale of (dark sector) inflation. The open parameter space is the same as Fig.~\ref{fig: Region 2} with an additional bound from the Trans-Planckian Censorship Conjecture (TCC), since the bounds from unwanted isocurvature perturbations (yellow region: $m<H_I$) and false vacuum stability considerations are both weaker than the upper bound on $\alpha$. The TCC bound closes the parameter space above it entirely at the ($m, \mu$) value shown in Tab.~\ref{tab: critical points}.}
 \label{fig: All region 2 Bounds}
 \end{figure}

Eqs.~\ref{eq: Delta V approx} and~\ref{eq: Vb approx} show that $\Delta V > V_{b}$ throughout Region 2, ensuring the tunneling field will indeed decay and not remain in the false vacuum forever. At this stage we will check that the barrier height is large enough to remain stable during inflation. Using these approximations, we find that all of Region 2 produces tunneling field potentials with barrier heights large enough to avoid destabilization from fluctuations caused by $T_{\rm{GH}}$, i.e. the Gibbons-Hawking bound is weaker than the upper limit on $\alpha$, as shown in Fig.~\ref{fig: All region 2 Bounds}.

Because $\mu\gg m$ throughout Region 2, there appears to be risk of isocurvature perturbations~\cite{kofman1987perturbations} being generated in the dark matter content (see~\cite{freese_dark_2023} for details). We find that throughout Region 2, $m > H_{I}$  where as before $H_{I} = \sqrt{\frac{\Delta V}{3 M_{\rm{p}}^2}}$ and $m$ is the mass of the tunneling field in the false vacuum. Therefore isocurvature perturbations are not induced since quantum fluctuations in $\phi$ are heavily suppressed~\cite{freese_dark_2023}. Fig.~\ref{fig: All region 2 Bounds} shows that imposing the bounds for false vacuum stability (Gibbons-Hawking) and the upper limit on the strength of the DBB (CMB $\Delta N_{\rm{eff}}$) ensure that isocurvature perturbations are safely avoided throughout Region 2.

 The most stringent bound on the energy scale of the phase transition is realized through the Trans-Planckian Censorship Conjecture (TCC)~\cite{bedroya2020trans}. This conjecture postulates an upper limit on the scale of inflation for a consistent theory of quantum gravity~\cite{bedroya2020TCCinflation,freese_dark_2023}

 \begin{equation}
     \Delta V \lesssim 10^{48} \rm{MeV}^{4}
     \label{eq: Planck Conj}
 \end{equation}
where $(\Delta V)^{1/4}$ is the energy scale of dark sector inflation. Imposing this bound closes the parameter space at much smaller $m$ and $\mu$ values than in Fig.~\ref{fig: Region 2}. The parameter space including limits on the scale of inflation from the TCC is shown in Fig.~\ref{fig: All region 2 Bounds}. 

Appendix~\ref{app: critical point eval} shows calculated PT and tunneling field parameters for the critical points in Tab.~\ref{tab: critical points}. A notable result of these calculations is that a DBB as late as $\approx$ 32.5 days (the latest a DBB can occur) after the Hot Big Bang can be realized in both Region 1 and Region 2. The only significant difference between these two regions (for a very late DBB) is the mass shift $m$ undergoes in transitioning from the false vacuum to the true vacuum. In both cases, the mass in the true vacuum $m_{\phi}$ is $\mathcal{O}(10^{-4}$ MeV) yet the starting masses (in the false vacuum) are very different. In Region 2, $m$ in the false vacuum is $\mathcal{O}(10^{-16}$ MeV) while in Region 1, $m$ is $\mathcal{O}(10^{-4}$ MeV) (see Tab.~\ref{tab:critical point analysis} in Appendix~\ref{app: critical point eval}).

\subsection{Benchmark Dark Big Bangs}\label{subsec: Benchmarks}
 A large range of dark matter particle masses $m_{\chi}$ can be realized through the DBB~\cite{freese_dark_2023}. The energy scale of dark sector inflation acts as a guide to the dark matter that can be considered. A dark sector with light dark matter requires $m_{\chi}\lesssim (\Delta V)^{1/4}$. On the other hand, dark matter can be more energetic than the energy scale of dark sector inflation by considering the Lorentz boost of the (runaway) bubble walls $\gamma_{w}$~\cite{freese_dark_2023, watkins_reheating_1992}. Heavy dark matter $m_{\chi} \gg (\Delta V)^{1/4}$ is then limited by~\cite{freese_dark_2023,watkins_reheating_1992}

 \begin{equation}
     m_{\chi} \lesssim \gamma_{w}m_{\phi} \simeq 2\frac{\Delta V}{m (\Delta\phi)^2}\frac{M_{P}}{\sqrt{g_{\rm{eff}}(T_{*})T_{*}^4}} m_{\phi}
     \label{eq: dark matter mass max}
 \end{equation}
 The maximum possible dark matter masses for the critical points of the parameter space are shown in Tab.~\ref{tab:critical point analysis}. The DBB is constrained to be consistent with the $\Lambda$CDM model of cosmology, so dark matter produced in the DBB must become cold early enough in the universe to reflect the $\Lambda$CDM picture of structure formation. The requirement that dark matter be non relativistic by $T > 7.8$ keV~\cite{freese_dark_2023} somewhat constrains choices of $m_{\chi}$ in constructing a model DBB. While there are many dark matter particle models that can be realized through a DBB~\cite{freese_dark_2023}, we will choose the simplest model: dark WIMPs. 
 
 Dark WIMPs are analogous to the well known WIMP model for particle dark matter, the difference being dark WIMPs annihilate into dark radiation $\xi$ that resides in the dark sector~\cite{freese_dark_2023}. We will choose both $\chi$ and $\xi$ to be real scalars for simplicity (following choices made in Ref.~\cite{freese_dark_2023}). The Lagrangian is then Eq.~\ref{eq:Dark Sector Lagrangian} with~\cite{freese_dark_2023}

 \begin{equation}
     \mathcal{L}_{\rm{DR}} \supset y'\chi^2\xi^2
     \label{eq: lagrangian, dark rad}
 \end{equation}
 where we have only included terms important to the annihilation process being considered. For our benchmark cases, we follow the standard thermal freeze out model using~\cite{freese_dark_2023}

\begin{equation}
    \frac{dY_{\chi}}{dT} = \frac{\langle \sigma_{\chi\chi\rightarrow\xi\xi}v\rangle s}{TH(T)}\left(1 + \frac{T}{3g_{\rm{eff}}}\frac{dg_{\rm{eff}}}{dT}\right)(Y_{\chi}^2-Y_{\chi,eq}^2)
    \label{eq:Dark Wimp Boltzmann}
\end{equation}
 as the Boltzmann equation for 
 $\chi\chi\leftrightarrow\xi\xi$ annihilations, where $Y_{\chi} = n_{\chi}/s$.
 
 The largest deviations from the freeze out of standard WIMPs come from the dark and visible sectors having different temperatures. In the Dark WIMP case, the equilibrium abundance $Y_{\chi,\rm{eq}}$ depends on $T_{\rm{DS}}$, as does when $\chi$ is considered to be relativistic $m_{\chi}/T_{\rm{DS}} < 1$. Since entropy is conserved in both sectors and, for the dark WIMP case, the dark sector always has relativistic degrees of freedom, the ratio of the temperatures of the two sectors is constant up to changes in the degrees of freedom~\cite{freese_dark_2023}

\begin{equation}
    \frac{T_{\rm{DS}}}{T}=\left(\frac{g_{\rm{eff}}(T)}{g_{\rm{eff}}(T_{*})}\right)^{1/3}\left(\frac{g_{\rm{DS}}(T_{\rm{DS,*}})}{g_{\rm{DS}}(T_{\rm{DS}})}\right)^{1/3}\frac{T_{\rm{DS,*}}}{T_{*}}
    \label{eq: Tds DW}
\end{equation}
Eq.~\ref{eq: Tds DW} is used to put $T_{\rm{DS}}$ in terms of $T$. For the benchmark cases considered here, we took $g_{\rm{DS},*} = 2$ if $\chi$ is produced relativistically and $g_{\rm{DS}} = 1$ once $\chi$ becomes non-relativistic ($\xi$ will always contribute to the dark sector degrees of freedom). 

To incorporate $g_{\rm{DS}}(T_{\rm{DS}})$ into our calculations, we created a continuous step function modeled after the behavior of $g_{\rm{eff}}$(T) in the visible sector. The step function goes from an initial value of the relativistic degrees of freedom in the dark sector $g_{\rm{DS},i}$ to a final value $g_{\rm{DS},f}$ with the transition centered at $T_{\rm{nr}}$. The steepness of the transition is determined by the parameter $K$, with a smaller $K$ giving a steeper transition. Therefore, we approximated $g_{\rm{DS}}$ as\footnote{Note that here, $g_{\rm{DS}}$ is a function of $T$, not $T_{\rm{DS}}$. This choice was made so Eq.~\ref{eq:gds(T)} can be used in Eq.~\ref{eq: Tds DW} to put $T_{\rm{DS}}$ in terms of $T$.}

\begin{equation}
    g_{\rm{DS}}(T)\approx \frac{g_{\rm{DS},f} - g_{\rm{DS},i}}{e^{(T - T_{\rm{nr}})/K} + 1} + g_{\rm{DS},i}
    \label{eq:gds(T)}
\end{equation}
 where $T_{\rm{nr}}$  is the visible sector temperature at which $\chi$ becomes non relativistic\footnote{We approximated this temperature by using Eq.~\ref{eq: Tds DW} when all of $\chi$ is non relativistic ($g_{\rm{DS}}(T_{\rm{DS}}) = 1$).} ($m_{\chi}/T_{\rm{DS}} = 1$) and we used $g_{\rm{DS},i} = 2$  and $g_{\rm{DS},f} = 1$ in the following analysis. For each Dark WIMP considered, we determined the steepness of the step function (the value of $K$) using two methods. The first method (columns 2 \& 4 in Tab.~\ref{tab: DBB Benchmarks}) chose $K$ such that 80\% of $\chi$ is non relativistic by $T_{\rm{DS}} \approx \frac{1}{6}m_{\chi}$~\cite{baumann_cosmo_2022}. A consequence of this is that $g_{\rm{DS},f} = 1$ is not quite reached by the universe today. Our second method for determining $K$ (columns 3 \& 5 in Tab.~\ref{tab: DBB Benchmarks}) chose $K$ such that $g_{\rm{DS},f} = 1$ is reached by $m_{\chi}/T_{\rm{DS}} \approx 10$. It is important to note that the different $g_{\rm{DS}}(T_{\rm{DS}})$ functions used only affect the relic abundance of $\chi$ slightly, and do not have an impact on the phase transition parameters which are entirely determined by $m,\mu$, and $\lambda$. The benchmark cases considered are shown in Tab.~\ref{tab: DBB Benchmarks}. The interaction strength ($y'$) has been adjusted so the correct relic abundance is achieved and we note that a smaller $K$ results in slightly cooler freeze out temperature ($T_{\rm{FO}}$). 

\begin{table}[!htb]
    \centering
    \begin{tabular}{|l|c|c|c|c|}\hline
     Benchmark & \multicolumn{2}{|c|}{DW 1}  & \multicolumn{2}{|c|}{DW 2} \\ \hline
     \multicolumn{5}{|c|}{Input Parameters} \\ \hline
     $m_{\chi}$ [keV]  & \multicolumn{2}{c|}{10} & \multicolumn{2}{c|}{500} \\ 
     $m$ [keV] & \multicolumn{2}{c|}{$8\times10^{-11}$} & \multicolumn{2}{c|}{$6 \times 10^{-5}$} \\
     $\mu$ [keV] & \multicolumn{2}{c|}{20} & \multicolumn{2}{c|}{$2 \times 10^5$} \\
     $\lambda$  & \multicolumn{2}{c|}{1} & \multicolumn{2}{c|}{1} \\
     $K$ & 0.012 & 0.001 & 0.34 & 0.04 \\
     $y'\times10^7$ & $0.0120$ & $0.0126$ & $0.189$ & $0.217$ \\
     \hline
     \multicolumn{5}{|c|}{Derived Parameters} \\ \hline
     $m_{\phi}$ [keV]  & \multicolumn{2}{c|}{30.0} & \multicolumn{2}{c|}{$3.00 \times 10^5$} \\
     $(\Delta V)^{1/4}$ [keV]  & \multicolumn{2}{c|}{11.4} & \multicolumn{2}{c|}{$1.14 \times 10^5$}\\
     \hline
     \multicolumn{5}{|c|}{Phase Transition} \\ \hline
     $t_{*}$ [s]  & \multicolumn{2}{c|}{1545} & \multicolumn{2}{c|}{$2.75 \times 10^{-5}$} \\
     $T_{*}$ [MeV]  & \multicolumn{2}{c|}{0.029} & \multicolumn{2}{c|}{138.39}  \\
     $T_{\rm{DS,*}}$ [MeV] & \multicolumn{2}{c|}{0.013} & \multicolumn{2}{c|}{126.55} \\ 
     $\alpha$  & \multicolumn{2}{c|}{0.02} & \multicolumn{2}{c|}{0.07} \\    \hline
     \multicolumn{5}{|c|}{Dark Matter} \\ \hline
     $\langle \sigma_{\chi\chi\rightarrow\xi\xi}v\rangle$ [$\rm{cm^3 s^{-1}}$] & $1.3\times 10^{-26}$ & $1.5\times 10^{-26}$ & $1.3 \times 10^{-27}$ & $1.8 \times 10^{-27}$ \\ 
     $T_{\rm{FO}}$ [keV]  & 2.37 & 2.16 & 62.78 & 58.60 \\
     $T_{\rm{nr}}$ [keV]  & \multicolumn{2}{c|}{18.4} & \multicolumn{2}{c|}{553.32}\\
     $\Omega_{\chi}h^2$  & \multicolumn{2}{c|}{0.120} & \multicolumn{2}{c|}{0.120} \\
     $\Delta N_{\rm{eff}}$  & \multicolumn{2}{c|}{0.24} & \multicolumn{2}{c|}{0.41} \\
      \hline
     
    \end{tabular}
    \caption{Benchmark Dark Big Bangs producing Dark WIMP (DW) dark matter with input parameters from Region 2. The input parameters were chosen to feature phase transitions that occur at different times in the universe, are strong enough to fall within the sensitivity bounds of upcoming GW surveys (Fig.~\ref{fig: GW Projections}), and produce the correct relic abundance assuming Dark WIMPs make up all the dark matter in the universe. DW 1 and DW 2 have been split into two columns corresponding to the steepness $K$ of the $g_{\rm{DS}}$ function used (Eq.~\ref{eq:gds(T)}).  Benchmark cases featuring the same Dark WIMP masses with parameters chosen from Region 1 can be found in Ref.~\cite{freese_dark_2023}.}
    \label{tab: DBB Benchmarks}
\end{table}

\begin{figure}[!htb]
 \includegraphics[width = 0.7\textwidth]{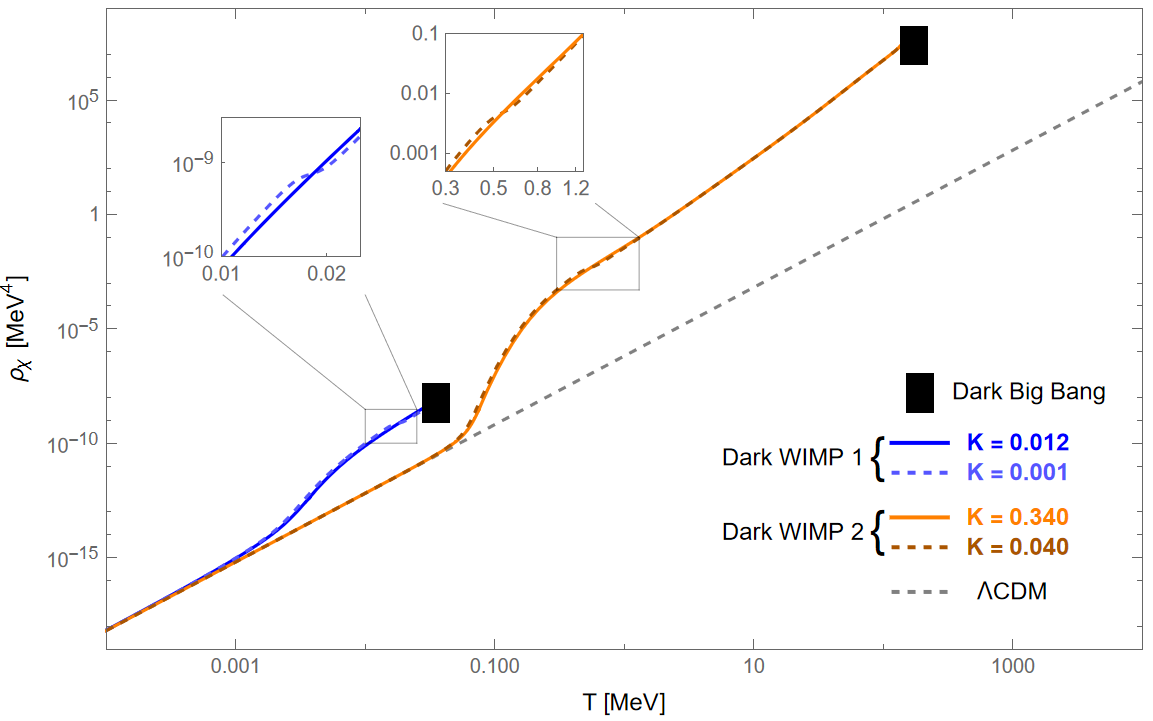}
 \caption{Evolution of the Dark WIMP energy density for the benchmark cases shown in Tab.~\ref{tab: DBB Benchmarks}. In all cases, the dark matter is produced relativistically and redshifts like $\rho_{\chi} \propto T^4$. The Dark WIMPs freeze out as cold relics and join the standard $\Lambda$CDM evolution track, redshifting like $\rho_{\chi} \propto T^3$, to produce the correct relic abundance of dark matter in the universe. Constructing $g_{\rm{DS}}(T)$ such that $g_{DS} = 1$ by $m_{\chi}/T_{\rm{DS}} \approx 10$ (the $K$ = 0.001, 0.040 dashed lines) results in a small bump in the energy density before Boltzmann suppression sets in, which has been highlighted in the zoomed regions. In both cases, the correct relic abundance is obtained by slightly increasing the interaction strength. The bump also causes an earlier $T_{\rm{FO}}$ (see Tab.~\ref{tab: DBB Benchmarks}).}
 \label{fig: Evolution}
 \end{figure}

The evolution of $\rho_{\chi}$ for each benchmark case is shown in Fig.~\ref{fig: Evolution}. Once $Y_{\chi}$ is solved for numerically, the dark WIMP energy density is given by $\rho_{\chi} = m_{\chi}Y_{\chi}s$ in the non relativistic regime. The dark WIMP energy density evolution while the particles are in thermal equilibrium and (for the most part) relativistic is given by:

\begin{equation}
    \rho_{\chi,\rm{eq}} = \frac{g_{\rm{DS}}(T)}{2\pi^2}m_{\chi}^3 T_{\rm{DS}} K_{1}\left(\frac{m_{\chi}}{T_{\rm{DS}}}\right) + T_{\rm{DS}}Y_{\chi,\rm{eq}} s 
    \label{eq: Equilibrium evolution}
\end{equation}
where $K_{1}$ is a modified Bessel function of the second kind and $s$ is the entropy density of the visible sector $s = \frac{2\pi^2}{45} g_{\rm{s}}(T) T^3$. Eq.~\ref{eq: Equilibrium evolution} is found by numerically integrating the Maxwell-Boltzmann phase space distributions for $n_{\chi,\rm{eq}}$ and $\rho_{\chi,\rm{eq}}$.\footnote{As in Ref.~\cite{freese_dark_2023}, we have approximated the Bose-Einstein distribution with a Maxwell-Boltzmann distribution.}

\begin{figure}[!htb]
 \includegraphics[width = 0.7\textwidth]{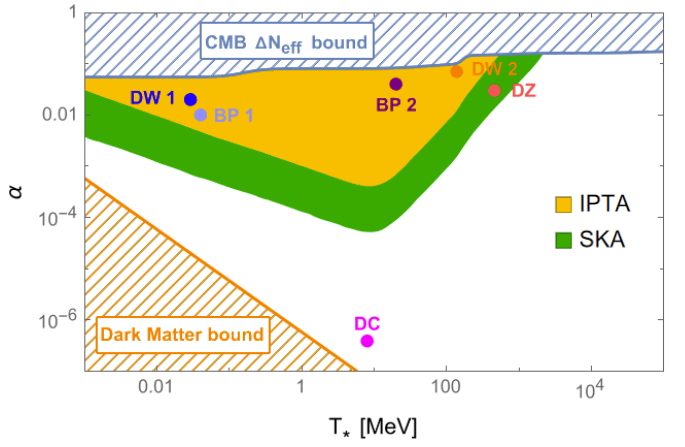}
 \caption{Benchmark DBB phase transition parameters on the projected sensitivity range (colored regions) of the upcoming IPTA and SKA pulsar timing arrays~\cite{freese_dark_2023}. The hatched regions correspond to the bounds on the DBB discussed in this paper and shown in Fig.~\ref{fig:DBB Bounds}. BP 1, BP 2, DZ, and DC, are the benchmarks considered in Ref.~\cite{freese_dark_2023}. Here we have used the $\alpha$ and $T_{*}$ values published in their tables, ignoring the discrepancies mentioned in Fig.~\ref{fig: Sliver Zoom 4} since the values tabulated can be achieved with slightly different choices for $m$ and $\mu$. To the figure we have added our new Dark WIMP benchmarks from Region 2: DW 1 and DW 2.}
 \label{fig: GW Projections}
 \end{figure}

 Before the DBB, all of the energy in the dark sector resides in the tunneling field $\phi$. After the DBB, the energy is split between dark matter particles, dark radiation, and gravity waves. The energy density of the gravity waves produced during the DBB can be approximated by~\cite{caprini2007gravitational, freese_dark_2023}

 \begin{equation}
     \rho_{\rm{GW,*}}\sim 0.05\hspace{0.5mm} \alpha\hspace{0.5mm} \Delta V
     \label{eq: gravwaves}
 \end{equation}
at the time of the DBB. The energy density of gravity waves will subsequently redshift like $\rho_{\rm{GW}}\sim a^{-4}$~\cite{freese_dark_2023}. The projected sensitivity windows of the International Pulsar Timing Array (IPTA)~\cite{antoniadis2022IPTA} and the Square Kilometer Array (SKA)~\cite{dewdney2009SKA} make strong and late DBB's favorable for detection~\cite{freese_dark_2023}. Both benchmark Dark Big Bangs fall within IPTA and SKA sensitivity windows, as shown in Fig.~\ref{fig: GW Projections}.

\section{Conclusion and Future Directions}
\label{sec: Conclusion}

The Dark Big Bang is an exciting new theory for the evolution of the universe. As WIMPs continue to evade detection, it is becoming increasingly important to consider dark sectors that are strongly decoupled from the visible sector. We have conducted an extensive analysis of the available parameter space for the tunneling potential of a DBB that is consistent with cosmological observations. Notably, we have shown that a region of parameter space exists (Region 2) far from the singularity of the Euclidean action. This region of parameter space is especially useful because many of the PT parameters can be determined through analytic approximations.  

We have shown that nearly all of the important characteristics describing a DBB can be put in terms of the tunneling field parameters $m$, $\mu$, and $\lambda$. Even features of the gravity waves released from the PT, one of the most important observables of the DBB, can be determined from the tunneling field parameters (Eq.~\ref{eq: gravwaves}). Evidence for the existence of a stochastic background of gravity waves permeating the universe was first detected by NanoGrav in June 2023~\cite{nanograv_GWB_2023}. As research on gravity waves continues to advance, signatures of a DBB may become differentiable from the stochastic background. In fact, signatures of the Dark Big Bang may have already been observed in NanoGrav's 15 year data set~\cite{freese_dark_2023}. In the case of Dark WIMPs, evidence for a Dark Big Bang can be observed through a simultaneous GW and CMB signal~\cite{freese_dark_2023}. The dark radiation density present in the dark sector contributes to $\Delta N_{\rm{eff}}$, which can be measured in the upcoming CMB observations like CMB-S4~\cite{abazajian2019cmbs4} and the Simons Observatory~\cite{ade2019simons}.

In the future, it would be interesting to see how the projected sensitivities of upcoming gravity wave surveys translate onto the parameter space of the tunneling field potentials. Future research could also consider weak couplings between the dark and visible sectors through portals. These couplings could give rise to additional observable signatures of the DBB~\cite{freese_dark_2023} and new bounds on the tunneling field potential parameters. 

\section*{Acknowledgments}

  R.C. would like to thank Saiyang Zhang for useful discussions during the later stages of this project. C.I. acknowledges funding from Colgate University via the Research Council and the Picker Interdisciplinary Science Institute.

\appendix

\section{Analytic Behavior}
\label{append: Analytic Behavior}

In this section we discuss the analytic behavior of the phase transition parameters to clarify why the parameter space is expected to look as it does. We will separate the parameter space into two regimes: Region 1 ($\mu \gtrsim$ m) and Region 2 ($\mu \gg $ m).

\subsection{Region 1}
\label{subsec: region 1}

Both $\alpha$ and $T_{*}$ depend on the Euclidean action of the bounce solution. In the thin wall approximation, which is good enough to understand the general behavior

\begin{equation}
    S \approx \frac{\pi^2\mu^6}{24\lambda(\mu^2-2\lambda m^2)^3}
    \label{S thin wall}
\end{equation}

S has a singularity at $\mu^2 = 2\lambda m^2$, which is the driving factor for the behavior of $\alpha$ and $T_{*}$ for Dark Big Bangs where $m$ and $\mu$ are comparable in size. When $\mu \gtrsim \sqrt{2\lambda}m$, we are right above the singularity in this region of parameter space. Setting $\lambda = 1$, $\alpha$ scales like:

\begin{equation}
    \alpha \sim m^{-2}S^{-1}e^{S/2}\Delta V
    \label{alpha scaling}
\end{equation}
Here we can ignore the $\Delta V$ term, since the singularity of S drives the behavior. Above the singularity, $\alpha$ decreases rapidly from positive infinity. The scaling of $T_{*}$ is 

\begin{equation}
    T_{*} \sim (m^4S^2e^{-S})^{1/8}
    \label{T scaling}
\end{equation}
Above the singularity, $T_{*}$ initially increases from zero, dominated by $S^2$, then decreases like $e^{-S}$. These scaling show that, in Region 1, the phase transition parameters are extremely sensitive to the values of $m$ and $\mu$. Being near the singularity means there is a very narrow parameter space in which $\alpha$ and $T_{*}$ both fall within the correct bounds. 

\subsection{Region 2}
\label{subsec: region 2}
Another allowed parameter space for a DBB consistent with the imposed bounds opens up when $\mu \gg m$ for reasons that will be analytically motivated here. When $\mu \gg m$, we are far from the singularity in $S$. It is convenient to rewrite the PT parameters in terms of the variable $\frac{m^2}{\mu^2}$. In the following analysis, we set higher order terms of $\frac{m^2}{\mu^2}$ to zero and set $\lambda = 1$. To first-order approximation, the Euclidean action scales like

\begin{equation}
    S \sim 1 - 6\frac{m^2}{\mu^2}
    \label{eq: S first order}
\end{equation}
and the potential difference scales like
\begin{equation}
    \Delta V \sim \mu^4 (1 - \frac{4}{3}\frac{m^2}{\mu^2})
    \label{eq: DelV scaling}
\end{equation}

Using these approximations and Eq. \ref{alpha scaling}, it can be shown that for $\mu \gg m$,

\begin{equation}
    \alpha \sim \frac{\mu^6}{m^4}
    \label{eq: alpha scaling region 2}
\end{equation}
and using Eq. \ref{T scaling}
\begin{equation}
    T_{*} \sim \frac{m}{\mu^{1/2}}
    \label{eq: T scaling region 2}
\end{equation}

These scalings show that region 2 of the parameter space should be expected. Near the singularity (Region 1), $\alpha$ decreases from positive infinity, shooting quickly through the allowed parameter space. Then, if $m$ is fixed and $\mu$ increased, $\alpha$ will return to allowed values from below the lower bound.

We can also find the slope, in $m$-$\mu$ space, of the bounds on the DBB when $\mu \gg m$. Writing the lower bound on $\alpha$ (Dark Matter) as a function of $\mu$, we find

\begin{equation}
    \mu \sim m^{6/11}
    \label{eq: lower bound scaling}
\end{equation}
For the upper bound on $\alpha$ (CMB $\Delta N_{\rm{eff}}$), we find

\begin{equation}
    \mu \sim m^{2/3}
    \label{eq: upper bound scaling}
\end{equation}
These scaling relations are reflected in the parameter space where the border of the upper bound (blue region) on $\alpha$ has a steeper slope than the border of the lower bound (orange region; see Fig.~\ref{fig: Region 2} or Fig.~\ref{fig: All region 2 Bounds}). The steepest slope is the temperature (Lyman-$\alpha$) bound (green region) which has the scaling $\mu \sim m^2$, as can be seen from Eq.~\ref{eq: T scaling region 2}.

\section{Critical Point Evaluation}
\label{app: critical point eval}

The purpose of this section is to give numerical values for the tunneling field potentials and phase transition parameters for the critical points tabulated in Tab.~\ref{tab: critical points}. Since the critical points found are numerical approximations for the boundaries of the parameter space, some values calculated violate the bounds imposed on the DBB and some fall just within the parameter space. Tab.~\ref{tab:critical point analysis} is meant to give a general idea of the tunneling field and PT parameters near the critical points. For all calculations, we set $\lambda = 1$.

\begin{table}[!htb]

    \begin{tabular}{|l|c|c|c|c|c|}\hline
         & R2 Opens & R1 Opens & Regions Merge & PS Closes (TCC) & PS  Closes  \\ \hline $m$ [MeV]
         & $1.90\times10^{-16}$ & $2.3160\times10^{-4}$ & $6.10\times10^9$ & $1.65\times10^{12}$& $8.17\times10^{19}$\\ 
         $\mu$ [MeV] &
         $2.10\times10^{-4}$ & $3.5915\times10^{-4}$ & $1.64\times10^{10}$ & $2.70\times10^{12}$ & $2.18\times10^{20}$ \\ \hline
         ($\Delta V)^{1/4}$ [MeV] & $1.20\times10^{-4}$ & $1.20\times10^{-4}$ & $8.41\times10^{9}$ & $1.01\times10^{12}$ & $1.12\times10^{20}$ \\
         $\Delta\phi$ [MeV] & $1.58\times10^{-4}$ & $2.03\times10^{-4}$ & $1.15\times10^{10}$ & $1.60\times10^{12}$ & $1.53\times10^{20}$ \\ 
         ($V_{b})^{1/4}$ [MeV] & 
         $7.75\times10^{-5}$ & $7.55\times10^{-5}$ & $1.41\times10^{9}$ &
         $5.17\times10^{11}$ &
         $1.89\times10^{19}$ \\
         $m_{\phi}$ [MeV] & $3.15\times10^{-4}$ & $3.35\times10^{-4}$ & $2.22\times10^{10}$ &
         $2.74\times10^{12}$ &
         $2.94\times10^{20}$ \\ \hline
         $t_{*}$ [s] & $2.88\times10^{6}$ & $2.86\times10^{6}$ &
         $6.13\times10^{-31}$ &
         $4.62\times10^{-25}$ &
         $4.66\times10^{-41}$ \\
         $T_{*}$ [MeV] & $6.78\times10^{-4}$ & $6.81\times10^{-4}$ & 
         $6.19\times10^{14}$ &
         $7.13\times10^{11}$ &
         $7.09\times10^{19}$ \\
         $T_{\rm{DS,*}}$ [MeV] & $1.33\times10^{-4}$ & $1.32\times10^{-4}$ & $9.33\times10^{9}$ &
         $1.11\times10^{12}$ &
         $1.24\times10^{20}$ \\
         $\alpha$ & $8.77\times10^{-4}$ &
         $8.50\times10^{-4}$ & 
         $9.69\times10^{-22}$ &
         $0.113$ & 0.170 \\ \hline 
         $m_{\chi,\rm{max}}$ [GeV] & $7.94\times10^{28}$ & $4.04\times10^{16}$ & $1.69\times10^{8}$ & $6.16\times10^{17}$ & $2.24\times10^{18}$ \\ \hline
         
    \end{tabular}
    \caption{Tunneling potential and phase transition parameters for the critical points in the $m$-$\mu$ parameter space. To calculate $T_{\rm{DS},*}$, we fixed $g_{\rm{DS}} = 2$. We have also included the maximum mass of dark matter candidates produced in each DBB scenario, according to Eq.~\ref{eq: dark matter mass max}. For the critical point where Region 2 (R2) opens (column 2), $T_{*}$ is slightly below the Lyman-$\alpha$ bound. Notably, for both critical points where the parameter space (PS) closes (columns 5 \& 6), $T_{\rm{DS},*} > T_{*}$ yet $\alpha$ falls within the bounds (it is typical for the dark sector reheating temperature to be less than the visible sector temperature at the DBB). This is a result of our choice for $g_{\rm{DS}}$, which is much smaller than $g_{\rm{eff}}$ at high temperatures.}
    \label{tab:critical point analysis}
\end{table}

\clearpage
\bibliography{Bibliography}

\end{document}